\documentclass[journal]{IEEEtran}
\usepackage{amsmath,amsfonts}
\usepackage{algorithmic}
\usepackage{algorithm}
\usepackage{array}
\usepackage[caption=false,font=normalsize,labelfont=sf,textfont=sf]{subfig}
\usepackage{textcomp}
\usepackage{stfloats}
\usepackage{url}
\usepackage{verbatim}
\usepackage{graphicx}
\usepackage{cite}
\newcommand{\C}{\mathbb{C}}
\newcommand{\R}{\mathbb{R}}
\newcommand{\Z}{\mathbb{Z}}

\begin{document}

\title{Recursive Filters as \\ Linear Time-Invariant Systems}

\author{Jonathan H. Manton,~\IEEEmembership{Fellow,~IEEE}}

\maketitle

An example of a filter structure commonly used in digital signal processing is
\begin{equation} \label{eq:half}
  y[k] = \frac12 y[k-1] + x[k], \quad k \in \Z
\end{equation}
where $\Z$ denotes the integers. Textbooks call this a recursive filter or an infinite impulse-response (IIR) filter. In a statistical context, it is known as an autoregressive model. The input $x[k]$ is a doubly-infinite sequence of complex-valued numbers: $x[k] \in \C$ for $-\infty < k < \infty$. The output $y[k]$ is not unique unless we initialise the filter by choosing a value for $y[-1] \in \C$. We can then iterate forwards to determine $y[k]$ uniquely for $k \geq 0$ (e.g., $y[0] = \frac12 y[-1] + x[0]$, $y[1] = \frac12 y[0] + x[1]$, \dots) and we can iterate backwards to determine $y[k]$ uniquely for $k < -1$ (e.g., solving $y[-1] = \frac12 y[-2] + x[-1]$ gives us $y[-2]$). 

Since the output of a non-initialised recursive filter is not unique, there are theoretical complications to studying it as a linear time-invariant (LTI) system. If an impulse $x[k] = \delta[k]$ (where $\delta[0]=1$ and $\delta[k]=0$ for $k \neq 0$) is applied, we get an infinite number of impulse responses, one for each choice of $y[-1]$. Which is the true impulse response? If we apply a sinusoid, or more simply, a complex exponential $x[k] = e^{\jmath 2 \pi f k}$, we need not obtain a complex exponential of the same frequency $f$ at the output, so how can a frequency response be defined? Choosing a fixed initialisation (such as $y[-1]=0$) does not solve the problem because the initialised filter is not time-invariant and need not produce a sinusoid in response to a sinusoidal input.

Another surprise comes when taking the $z$-transform then rearranging: $Y(z) = 1 / (1 - \frac12 z^{-1}) \, X(z)$. If $X(z)=0$, we have only a single solution, $Y(z)=0$, so the $z$-transform has somehow discarded an infinite number of solutions. (For more interesting inputs, the $z$-transform may give us a few more solutions, depending on the region-of-convergence. We have still lost an infinite number of solutions.)

Textbooks do not draw attention to these issues because most of the time ``things just work''. This lecture note explains what is happening behind the scenes as well as the origin of the region-of-convergence concept in z-transform theory.

\section{Relevance}

Recursive filters are ubiquitous in signal processing because they are easy to implement. It is tempting to treat them as linear time-invariant systems since then a large theory can be brought to bear on their design and analysis, including Fourier analysis. Rather than cover up the gaps between recursive filters and true linear time-invariant systems, we explain directly how to manage the gaps. This results in a cleaner theory with no hidden surprises.

\section{Prerequisites and notation}

We assume the reader is comfortable with the basics of linear filters and their impulse and frequency responses. We have used the material presented here in an introductory course on digital signal processing at the undergraduate level.

Recall that $\delta[k]$ denotes the unit impulse $\delta[k]=0$ for $k \neq 0$ and $\delta[0]=1$. The unit step $u[k]$ is given by $u[k]=0$ for $k < 0$ and $u[k]=1$ for $k \geq 0$. The convolution $y[k] = \sum_{i=-\infty}^\infty h[i] x[k-i]$ of the sequences $h[k]$ and $x[k]$ is denoted $y = h \ast x$. Importantly, given a sequence $h[k]$, the convolution $y = h \ast x$ defines an LTI system with impulse response $h$. (Substituting $x[k] = \delta[k]$ into $y[k] = \sum_{i=-\infty}^\infty h[i] x[k-i]$ results in $y[k] = h[k]$.)

Use will be made of the Taylor series expansion
\begin{equation}
  (1-z)^{-1} = \sum_{k=0}^\infty z^k,\quad |z| < 1.
\end{equation}

\section{Problem statement}

Consider the recursive filter
\begin{equation} \label{eq:ar}
  y[k] = \sum_{i=1}^N \alpha_i \, y[k-i] + x[k],\quad k \in \Z
\end{equation}
where $\alpha_1,\cdots,\alpha_N \in \C$ are the filter coefficients. Practical implementations generally use $y[-1] = \cdots = y[-N]=0$ as the initialisation. Initialisation ensures that for any given input $x[k]$ there is a unique output $y[k]$ for $-\infty < k < \infty$. Technically though, such a system does not have a frequency response because the initialisation will, in general, prevent there being a solution of the form $y[k] = A e^{\jmath 2 \pi f k}$ when $x[k] = e^{\jmath 2 \pi f k}$. Here, $f \in \R$ is a fixed frequency and $A \in \C$ represents a possible change in the signal's amplitude and phase.

How can we rigorously apply the theory of linear time-invariant systems to recursive filters?

Separately, questions around the z-transform are answered. Why does it fail to find all the solutions of a recursive filter? Where does the concept of region-of-convergence come from?

\section{Solution}

Given the recursive filter \eqref{eq:ar}, define the sequence $h[k]$ to be the output $y[k]$ when the input is an impulse $x[k] = \delta[k]$ and when $y[k]=0$ for $k < 0$. While it is tempting to call this the impulse response of the filter, it is safest to think that \eqref{eq:ar} is not an LTI system and hence does not have an impulse response. More will be said about this presently.

We define the \textbf{associated LTI system} of the recursive filter \eqref{eq:ar} to be $\tilde y = h \ast x$. The impulse response of this LTI system is $h$. (Indeed, if $x[k] = \delta[k]$ then $y = h \ast x = h \ast \delta = h$.)

Importantly, there are three distinct systems under consideration. There is the non-initialised filter \eqref{eq:ar}. For any given input, there are an infinite number of outputs (unless all the filter coefficients are zero). If we understand an LTI system as having a unique output for a given input then this non-initialised filter is not LTI. Combining \eqref{eq:ar} with initial values for $y[-N],\cdots,y[-1]$ gives an initialised filter. We cannot call this an LTI system either. Unless all the initial values are zero then the system is not linear. Even if it was zero-initialised, it is not time invariant. (If the zero-initialised system were time invariant then shifting $x$ to the left one place should result in $y$ being shifted one place to the left, but $y[-1]$ is always stuck at 0.) The third system is our associated LTI system $\tilde y = h \ast x$. By construction, it is LTI.

The full theory of LTI systems can be applied to $\tilde y = h \ast x$. Although the recursive filter differs from this associated LTI system, we will see that they share many properties. It is often possible to work with the associated LTI system without explicitly computing $h$. This makes it appear we are directly applying LTI techniques to recursive filters.

The first result is that the associated LTI system picks out a particular solution of the non-initialised filter equation.

\noindent\textbf{Fact 1:}
Assume $x[k]$ and $\tilde y[k]$ satisfy $\tilde y = h \ast x$. Then $x[k]$ and $y[k] = \tilde y[k]$ satisfy the recursive filter equation \eqref{eq:ar}.

While a proof is given in the appendix, the following intuition is perhaps more informative. While we do not wish to claim \eqref{eq:ar} is an LTI system because it has an infinite number of solutions for any input, it nevertheless satisfies the following generalised LTI properties. Let $(x_1[k], y_1[k])$ and $(x_2[k], y_2[k])$ be two solution pairs. (We say $(x_1[k], y_1[k])$ is a solution pair if \eqref{eq:ar} is satisfied with $x[k]=x_1[k]$ and $y[k] = y_1[k]$.) Then for any scalars $a,b \in \C$ we have that $(a x_1[k] + b x_2[k], a y_1[k] + b y_2[k])$ is also a solution pair. Furthermore, for any integer $m$, define $x_3[k] = x_1[k-m]$ and $y_3[k] = y_1[k-m]$. Then $(x_3[k], y_3[k])$ is also a solution pair. For brevity, we will also express this by saying $(x_1[k-m], y_1[k-m])$ is a solution pair.

Now, the input-output pair $(\delta[k], h[k])$ satisfies \eqref{eq:ar} by definition of $h[k]$. By time invariance, $(\delta[k-i], h[k-i])$ is also a solution pair for any integer $i$. By linearity, any finite sum $(\sum_i x[i] \delta[k-i], \sum_i x[i] h[k-i])$ is a solution pair. If the sequence $x[k]$ only has a finite number of non-zero terms then we can write these summations as infinite summations. (Alternatively, we can assume the recursive filter is stable and appeal to continuity~\cite{zhangImportanceContinuityLinear2020}.) Let $\bar x[k] = \sum_i x[i] \delta[k-i]$ and $\bar y[k] = \sum_i x[i] h[k-i]$. Then $\bar x[k] = x[k]$ and $\bar y = x \ast h = h \ast x = \tilde y$, showing that our associated LTI system produces a solution pair for the recursive filter.

\noindent\textbf{Fact 2:}
Assume $x[k]$ is causal, meaning $x[k] = 0$ for $k < 0$. Let $y[k]$ be the output of the zero-initialised recursive filter \eqref{eq:ar}. Let $\tilde y[k]$ be the output of the associated LTI system: $\tilde y = h \ast x$. Then $y[k] = \tilde y[k]$ for all $k \in \Z$.

To verify Fact 2, first recall that the initialisation determines the output uniquely: iterating forwards from $k=0$ gives $y[k]$ for $k \geq 0$ and iterating backwards from $k=-1$ gives $y[k]$ for $k < -N$. Fact 1 tells us that $(x[k], \tilde y[k])$ is a solution pair of the general recursive filter. It will therefore be the output of the initialised recursive filter if and only if $\tilde y[k]$ satisfies the initial conditions, namely, $\tilde y[-N] = \cdots = \tilde y[-1] = 0$. Since $x[k]$ is causal by assumption and $h[k]$ is causal by construction, we know that $\tilde y[k]$ is causal; this is a standard property of LTI systems and follows immediately from the definition of convolution. This implies $\tilde y[-N] = \cdots = \tilde y[-1] =0$, proving Fact 2.

Generally in practice, the initialised filter equations are only applied in the forward direction: a singly-infinite sequence $x[k]$ for $k \geq 0$ is mapped to a singly-infinite sequence $y[k]$ for $k \geq 0$. An alternative interpretation of Fact 2 is that the associated LTI system is the \textbf{unique extension of this zero-initialised recursive filter to doubly-infinite sequences} such that the extension is an LTI system and the two systems agree on causal signals.

Before explaining how Fourier analysis of the associated LTI system provides insight into the initialised recursive filter, first we recall briefly the core tenet of Fourier analysis: sinusoids are eigenfunctions of LTI systems. If $x[k] = e^{\jmath 2 \pi f k}$ then
\begin{align}
  \tilde y[k] &= \sum_{i=-\infty}^\infty h[i] e^{\jmath 2 \pi f (k-i)} \\
    &= \left( \sum_{i=-\infty}^\infty h[i] e^{-\jmath 2 \pi f i} \right) e^{\jmath 2 \pi f k} \\
    &= H(f) e^{\jmath 2 \pi f k}
\end{align}
where $H(f) = \sum_{i=-\infty}^\infty h[i] e^{-\jmath 2 \pi f i}$ is the discrete-time Fourier transform of the impulse response $h$ of the LTI system $\tilde y = h \ast x$. Besides an amplitude and phase shift determined by $H(f)$, the ``sinusoid'' $e^{\jmath 2 \pi f k}$ passes through unaltered. The same is not true of our recursive filter when it incorporates initial values.

\noindent\textbf{Fact 3:}
Fix an input $x[k]$ and let $\tilde y[k]$ be the output of the associated LTI system $\tilde y = h \ast x$. The pair $(x[k], \tilde y[k] + y_0[k])$ is a solution pair of the recursive filter \eqref{eq:ar} if and only if $y_0[k]$ is a \textbf{homogeneous solution}, meaning it satisfies \eqref{eq:ar} with $x[k]=0$ for all $k$.

Fact 3 follows from Fact 1 and linearity. Since $(x,\tilde y)$ is a solution pair, $(x,y)$ is a solution pair if and only if $(x-x, \tilde y - y)$ is a solution pair, in other words, if and only if $\tilde y - y$ is a homogeneous solution.

Fact 3 allows us to understand precisely the difference between the output of our initialised recursive filter $y$ and our associated LTI system $\tilde y$. To illustrate, assume $x[k] = e^{\jmath 2 \pi f k}$. From above, $\tilde y[k] = H(f) e^{\jmath 2 \pi f k}$. From Fact 3, $\tilde y - y$ is a homogeneous solution. Like any solution, it is uniquely determined by its values at times $k=-N,\cdots,-1$. Had we initialised our filter using $y[-N] = \tilde y[-N],\cdots,y[-1]=\tilde y[-1]$, we would have had $y[k] = \tilde y[k]$ for all $k$. Otherwise, the difference between the two systems is due to the different initialisations.

Normally the recursive filter is designed to be stable, which means any homogeneous solution decays to zero as $k \rightarrow \infty$. For stable filters, the effect of the initial condition is eventually forgotten~\cite{engelbergAdvantagesForgeteryLecture2015}.

For perfectionists thinking this difference means Fourier analysis of our associated system only tells us our \emph{asymptotic} filter behaviour, this is misleading. In the real world, signals have finite duration. A filter that works for perfect sinusoids but fails for windowed sinusoids will not be useful. Therefore, regardless of the type of filter, we should be concerned with how fast it reacts to change. One measure is inputting $u[k] e^{\jmath 2 \pi f k}$ and seeing how long it takes the output to become sufficiently close to $H(f) e^{\jmath 2 \pi f k}$. Because the input $u[k] e^{\jmath 2 \pi f k}$ is causal, the associated LTI system will have the identical output to our recursive filter initialised to zero. The associated LTI system gives us the exact output, not just the asymptotic output. Both systems have the same ``reaction time''. It is easy to point to the exact cause of this reaction time for recursive filters, but all filters will have a reaction time.

\noindent\textbf{Fact 4:}
We can find the frequency response $H(f) = \sum_{i=-\infty}^\infty h[i] e^{-\jmath 2 \pi f i}$ of the associated LTI system directly from \eqref{eq:ar} by substituting $x[k] = e^{-\jmath 2 \pi f k}$ and $y[k] = H(f) e^{-\jmath 2 \pi f k}$ into \eqref{eq:ar} and solving for $H(f)$. If the filter is stable then the solution will be unique. (This is equivalent to the standard method of using the z-transform to find the frequency response.)

That there is at least one solution of the form $y[k] = H(f) e^{-\jmath 2 \pi f k}$ follows from Fact 1. There can only be another solution of this form if $e^{-\jmath 2 \pi f k}$ is a homogeneous solution. If the filter is stable though, all homogeneous solutions decay to zero. (The magnitude of $e^{-\jmath 2 \pi f k}$ is unity for all $k$.)

\section{Numerical example}

To find the associated LTI system of \eqref{eq:half}, start by replacing $x[k]$ by $\delta[k]$. We are interested in the solution $y[k]$ satisfying $y[k]=0$ for $k < 0$. Note that $y[k] = \frac12 y[k-1]$ is consistent with $y[k]=0$ for $k < 0$. For $k=0$ we have $y[0] = \frac12 y[-1] + 1 = 1$. Continuing, we have $y[1] = \frac12 y[0] + 0 = \frac12$ and $y[2] = \frac12 y[1] + 0 = (\frac12)^2$. In general, $y[k] = (\frac12)^k$ for $k \geq 0$. The associated LTI system is therefore $\tilde y = h \ast x$ where $h[k] = (\frac12)^k u[k]$ for $k \in \Z$.

The zero-initialised filter \eqref{eq:half} and the associated LTI system are different. Take the shifted impulse $x[-1] = 1$ and $x[k] = 0$ for $k \neq -1$. The output of \eqref{eq:half} with initial condition $y[-1]=0$ obviously satisfies $y[-1]=0$. (It also has $y[0] = 0$ and $y[-2] = 2 y[-1] - 2 x[-1] = -2$.) The output of the LTI system $\tilde y = h \ast x$ satisfies $\tilde y[k] = \sum_{i=-\infty}^\infty (\frac12)^i u[i] x[k-i] = (\frac12)^{k+1} u[k+1]$. In particular, $\tilde y[-1] = 1 \neq y[-1]$.

For causal signals the outputs will be the same (Fact 2). For example, if $x[k]$ is zero except for $x[0]=1$ and $x[1]=2$ then the output of the associated LTI system is the convolution of $\{1,2\}$ with the causal impulse response $\{1,\frac12,(\frac12)^2,\cdots\}$, namely, $\{1, 2 \frac12, 1 \frac14, \cdots \}$. The first three outputs of the zero-initialised recursive filter are
\begin{align}
  y[0] &= \frac12 y[-1] + x[0] = 1, \\
  y[1] &= \frac12 y[0] + x[1] = 2 \frac12, \\
  y[2] &= \frac12 y[1] + x[2] = 1 \frac14.
\end{align}

The frequency response $H(f)$ of the associated LTI system is the discrete-time Fourier transform of the impulse response $h[k] = (\frac12)^k u[k]$.
\begin{align}
  H(f) &= \sum_{k=-\infty}^\infty h[k] e^{-\jmath 2 \pi f k} \\
    &=\sum_{k=0}^\infty \left(\frac12\right)^k e^{-\jmath 2 \pi f k} \\
    &= \sum_{k=0}^\infty \left(\frac12 e^{-\jmath 2 \pi f} \right)^k \\
    &= \frac1{1 - \frac12 e^{-\jmath 2 \pi f}}. \label{eq:Hf}
\end{align}

The easier way of finding the frequency response is to apply Fact 4. Substituting $x[k] = e^{\jmath 2 \pi f k}$ and $y[k] = H(f) e^{\jmath 2 \pi f k}$ into \eqref{eq:half} yields
\begin{align}
  H(f) e^{\jmath 2 \pi f k} &= \frac12 H(f) e^{\jmath 2 \pi f (k-1)} + e^{\jmath 2 \pi f k} \\
    &= \frac12 H(f) e^{-\jmath 2 \pi f} e^{\jmath 2 \pi f k} + e^{\jmath 2 \pi f k}.
\end{align}
Dividing through by $e^{\jmath 2 \pi f k}$ shows $H(f) = \frac12 H(f) e^{-\jmath 2 \pi f} + 1$, agreeing with \eqref{eq:Hf}. (The equivalent calculation is performed by taking the z-transform then substituting $z=e^{\jmath 2 \pi f}$. Indeed, $Y(z) = \frac12 z^{-1} Y(z) + X(z)$ shows $H(z) = Y(z) / X(z) = 1 / (1 - \frac12 z^{-1})$.)

In practice, applying $x[k] = e^{\jmath 2 \pi f k}$ to \eqref{eq:half} means initialising the recursive filter to $y[-1]=0$ then generating $y[k]$ for $k \geq 0$ recursively. Due to this initialisation, the output will not agree with $H(f) e^{\jmath 2 \pi f k}$ but it will converge to it. If we wanted the recursive filter to produce $H(f) e^{\jmath 2 \pi f k}$ then we would need to initialise it with $y[-1] = H(f) e^{\jmath 2 \pi f (-1)}$. This requires a different initialisation for each frequency $f$ and thus is only of theoretical interest.

If we wanted our associated LTI system to yield the same output as our zero-initialised recursive filter in response to $x[k] = e^{\jmath 2 \pi f k}$ then we simply recognise that we will get the same output from our recursive filter for non-negative times whether $x[k] = e^{\jmath 2 \pi f k}$ or $x[k] = e^{\jmath 2 \pi f k} u[k]$ is the input. By using the latter signal, which is causal, our associated LTI system will produce the same output as our zero-initialised recursive filter (Fact 2).

To see why the output $y[k]$ of the recursive filter will converge to $H(f) e^{\jmath 2 \pi f k}$ if $x[k] = e^{\jmath 2 \pi f k}$ is the input, recall that $y[k] - H(f) e^{\jmath 2 \pi f k}$ must be a homogeneous solution.
All homogeneous solutions $y_0[k]$ of \eqref{eq:half} are of the form $y_0[k] = c (\frac12)^k$ for an arbitrary constant $c \in \C$. Indeed, we are free to choose $y_0[0]$ as we like, then $y_0[k] = \frac12 y_0[k-1]$ forces $y_0[k] = (\frac12)^k y_0[0]$ (as can be proved by induction). The difference between the actual initialisation $y[-1]$ and $H(f) e^{\jmath 2 \pi f (-1)}$ determines the constant $c$. For all $k$ we then have $y[k] - H(f) e^{\jmath 2 \pi f k} = c (\frac12)^k$ which converges rapidly to zero as $k \rightarrow \infty$.

\section{Region of convergence} \label{sec:roc}

The recursive filter \eqref{eq:ar} is known to mathematicians as a finite-difference equation. If we are only interested in iterating forwards then the theory of generating functions can be used to find all solutions. The z-transform essentially mimics the generating function approach, yet it fails to find all solutions. This section investigates why. In a few places, mention is made of some deeper mathematical concepts which can be ignored if so desired.

It suffices to consider the special case \eqref{eq:half}. Naively mimicking the generating-function approach leads to the following flawed derivation. To each sequence $\{ y[k] \}_{k=-\infty}^\infty$ we associate a ``formal power series'' $\cdots + y[-1] z^{1} + y[0] + y[1] z^{-1} + y[2] z^{-2} + \cdots$. We simply use powers of $z$ to separate the different terms of our sequence. (Engineers prefer the terms to be $y[k] z^{-k}$ rather than the more obvious $y[k] z^k$. It makes little difference.) In engineering parlance, we form the $z$-transform $Y(z) = \sum_{k=-\infty}^\infty y[k] z^{-k}$. Observe what happens when we write $Y(z) = \frac12 z^{-1} Y(z) + X(z)$.
\begin{multline}
  \cdots + y[-1] z + y[0] + y[1] z^{-1} + \cdots \\
  = \cdots + \frac12 y[-2] z + \frac12 y[-1] + \frac12 y[0] z^{-1} + \cdots \\
  + \cdots + x[-1] z + x[0] + x[1] z^{-1} + \cdots.
\end{multline}
Equating the coefficients of the powers of $z$ shown above, we get the equations $y[-1] = \frac12 y[-2] + x[-1]$, $y[0] = \frac12 y[-1] + x[0]$ and $y[1] = \frac12 y[0] + x[1]$. If we continue this for all powers of $z$, we see we have recreated \eqref{eq:half} precisely.

We would therefore expect a one-to-one correspondence between every solution of
\begin{equation} \label{eq:z1}
  \left(1 - \frac12 z^{-1}\right) Y(z) = X(z)
\end{equation}
and every solution of \eqref{eq:half}. Going further, we would expect the same to be true of
\begin{equation} \label{eq:z2}
  Y(z) = \left(1 - \frac12 z^{-1}\right)^{-1} X(z).
\end{equation}
Naively, this suggests there is only a single solution to \eqref{eq:half} for a given input. Readers familiar with $z$-transform theory may recall the concept of region of convergence and realise we can sometimes extract multiple solutions. Nevertheless, despite our careful argument above, we have lost an infinite number of solutions. How careless!

We know \eqref{eq:half} has non-trivial homogeneous solutions, such as $y[k] = (\frac12)^k$. This corresponds to
\begin{equation} \label{eq:Yzhom}
  Y(z) = \cdots + 4 z^2 + 2 z + 1 + (1/2) z^{-1} + (1/4) z^{-2} + \cdots.
\end{equation}
Direct evaluation term-by-term shows $Y(z) - \frac12 z^{-1} Y(z) = 0$. In \eqref{eq:z1}, this means we have the product of two non-zero functions, $1-\frac12 z^{-1}$ and $Y(z)$, equalling zero! Our ``formal power series'' misbehaves algebraically.

Actually, double-sided formal power series do not exist in that multiplication cannot be defined sensibly. Multiplication of single-sided formal power series can be defined term-by-term because each term in the product involves only a finite number of coefficients from both series. Attempting to multiply two double-sided power series involves infinite summations, meaning convergence becomes an issue; we cannot work ``formally''. The closest we can come is using formal Laurent series, which would limit us to sequences that are zero for all distant past (or all distant future).

The only way to take this approach further is to work with actual functions and not formal functions. Two functions can be multiplied pointwise: if $C(z) = A(z) B(z)$ then to compute $C(z)$ for a specific $z$ we simply evaluate $A(z)$ and $B(z)$ at that point then multiply the resulting numbers together. What this means though is that all the functions in \eqref{eq:z1} must be able to be evaluated at (sufficiently many) points $z \in \C$.

The function $1-\frac12 z^{-1}$ can be evaluated everywhere except at $z=0$. So \eqref{eq:z1} tells us that if $X(z)=0$ then $Y(z)=0$ for $z \neq 0$. We have still failed to find the solution \eqref{eq:Yzhom}. The reason though is that we cannot evaluate \eqref{eq:Yzhom} at any point $z$. \textbf{By restricting attention to well-defined doubly-infinite power series, we have precluded some sequences from having a corresponding power series.} (For infinite summations to behave well, we require them to be absolutely summable. If $|z| \geq 1$ then the portion $1+2z+4z^2+\cdots$ blows up, while if $|z|<1$ then the portion $1+\frac12 z^{-1} + \frac14z^{-2}+\cdots$ blows up.)

We require $X(z)$ and $Y(z)$ to be more than just arbitrary functions: to recover $y[k]$ from $Y(z)$ requires a way of expressing $y[k]$ as a doubly-infinite power series. Even for single-sided power series there are difficulties because not every function equals its Taylor series in a sufficiently small interval. This suggests restricting attention to analytic functions: an analytic function is a function expressible locally as a convergent power series.

Given an analytic function $f$ defined in a neighbourhood of a point $a$, we can expand it in a power series about $a$, meaning we can write $f(z) = \sum_{k=0}^\infty a_k (z-a)^k$ for $|z-a|$ sufficiently small. If $|z-a|$ is too large then $\sum_{k=0}^\infty a_k (z-a)^k$ need not exist, even if $f(z)$ does. Indeed, any power series will either exist for all $z$, or there will be a constant $R$ such that it converges for $|z-a| < R$ and diverges for $|z-a| > R$. This $R$ is called the radius of convergence.

For recursive filters, at first it appears that the only $a \in \C$ of interest is $a=0$ because we want to find the coefficients of $z^k$. However, this only gives us half the $y[k]$ terms because our power series is single sided. Since $\sum_{k=-\infty}^0 a_k z^k = \sum_{k=0}^\infty a_{-k} (1/z)^k$, an obvious trick is to replace $z$ by $z^{-1}$. Mathematically, this is equivalent to extending the complex plane to include a point at infinity, thereby obtaining the Riemann sphere, then expanding about the point at infinity ($a=\infty$). For recursive filters then, we are interested in expansions of $f$ about either $a=0$ or $a=\infty$.

For concreteness, consider $Y(z) = (1-z)^{-1}$. Expanding about $a=0$ gives $Y(z) = \sum_{k=0}^\infty z^k$, corresponding to the sequence $y[k] = u[-k]$. Expanding about $a=\infty$ gives $Y(z) = -\sum_{k=1}^\infty z^{-k}$, corresponding to the sequence $y[k] = \delta[k] - u[k]$. (Since $Y(z^{-1}) = (1-z^{-1})^{-1} = z/(z-1) = (-z)(1-z)^{-1}$, we have $Y(z^{-1}) = (-z) \sum_{k=0}^\infty z^k$, and so $Y(z) = (-z^{-1}) \sum_{k=0}^\infty z^{-k} = -\sum_{k=1}^\infty z^{-k}$.) The reason for obtaining two different series from the one function is that the expansion was performed about two different points ($a=0$ and $a=\infty$). This manifests itself in the region of convergence (ROC) of the two series. The ROC must include the point about which the series was computed, and the theory tells us the ROC will always be of the form $|z-a| < R$. In the case of $a=\infty$ this translates to a ROC of the form $|z| > R$ (as can be seen by replacing $z$ with $z^{-1}$ and $a=\infty$ with $a=1/\infty=0$). For $Y(z) = \sum_{k=0}^\infty z^k$, the ROC is $|z|<1$. (If $z=1$ then the series diverges.) For $Y(z) = -\sum_{k=1}^\infty z^{-k}$, the ROC is $|z|>1$. (Again, if $z=1$ then the series diverges.)

The ROC is crucial for equations such as \eqref{eq:z2} to make sense: unless there is a common ROC amongst the functions $X(z)$, $Y(z)$ and $H(z)$, we cannot write $Y(z) = H(z) X(z)$, at least not when we are using the power series representations of these functions.

Given a sequence $x[k]$, either $\sum_{k=-\infty}^\infty x[k] z^{-k}$ does not converge in any open region, in which case the sequence fails to have a $z$-transform, or it has a region of convergence of the form $r < |z| < R$ (including the special cases $|z| < R$, $|z| > r$ and $z \in \C$). Every sequence has a unique $z$-transform assuming it exists. The only subtlety is if we replace $\sum_{k=-\infty}^\infty x[k] z^{-k}$ by an explicit function then the function may be defined on a larger domain than the original ROC, which is why it is important to keep track of the ROC. For example, while $\sum_{k=0}^\infty z^k$ is only defined on $|z|<1$, its analytic continuation $(1-z)^{-1}$ is defined everywhere except $z=1$. Losing track of the ROC would mean not knowing whether to expand $(1-z)^{-1}$ about $0$ or $\infty$.

Interestingly, when taking the $z$-transform of a recursive filter, we obtain the transfer function $H(z)$ as a rational function but without a ROC, as exemplified in \eqref{eq:z2} with $H(z) = (1-\frac12z^{-1})^{-1}$. If $H(z)$ has $n$ poles with distinct magnitudes $|p_1| < \cdots < |p_n|$ then there are $n+1$ possible ROCs: $|z| < |p_1|$, $|p_1| < |z| < |p_2|$, $\cdots$, $|p_{n-1}| < |z| < |p_n|$, $|p_n| < |z|$. Each one will lead to a valid solution pair for the recursive filter \eqref{eq:ar} (provided $X(z)$ is well-defined on the chosen ROC). From the infinite number of solution pairs, the $z$-transform approach finds a finite number. The other solutions correspond to sequences not having valid $z$-transforms because the corresponding power series diverge to infinity. (Intuitively, this is because almost all homogeneous solutions will be unbounded in both directions.)

Since the $z$-transform of $\delta[k]$ is $X(z)=1$, each choice of ROC for $H(z)$ can be thought of as the $z$-transform of an impulse response of \eqref{eq:ar}. (Any $h[k]$ for which $(\delta[k],h[k])$ is a solution pair of \eqref{eq:ar} can be called an impulse response; there are infinitely many.) The associated LTI system is formed from the impulse response which is causal.

\noindent\textbf{Fact 5:} The transfer function $H(z)$ of the recursive filter \eqref{eq:ar} is the $z$-transform of the impulse response $h[k]$ of the associated LTI system where the corresponding ROC is of the form $r < |z|$. (Technically, $r$ is the largest magnitude of the poles of $H(z)$.)

By definition, $h[k]$ is the unique solution pair $(\delta[k],h[k])$ with $h[k]=0$ for $k<0$. Let $\tilde h[k]$ be the inverse $z$-transform of $H(z)$ using the ROC $r < |z|$. This means taking the power series about $\infty$, thereby writing $H(z) = \sum_{k=0}^\infty \tilde h[k] z^{-k}$. Setting $\tilde h[k]=0$ for $k<0$ allows us to write $H(z) = \sum_{k=-\infty}^\infty \tilde h[k] z^{-k}$. The $z$-transform of $x[k]=\delta[k]$ is $X(z)=1$. Since $Y(z)=H(z)X(z)$, $(\delta[k], \tilde h[k])$ is also a solution pair, and by the aforementioned uniqueness, we must have $h[k] = \tilde h[k]$, proving Fact 5.

We conclude with an example of how the above the theory brings clarity. Consider the following system as a special case of \eqref{eq:ar}.
\begin{equation}
  y[k] = 2\frac12 y[k-1] - y[k-2] + x[k].
\end{equation}
Taking the z-transform shows $Y(z) = H(z) X(z)$ where
\begin{equation}
  H(z) = \left( 1 - 2\frac12 z^{-1} + z^{-2}\right)^{-1}.
\end{equation}
Substituting $z = e^{\jmath 2 \pi f}$ gives a well-defined frequency response $H(e^{\jmath 2 \pi f})$. On the other hand, we can compute the associated LTI system by solving $h[k] = 2\frac12 h[k-1] - h[k-2] + \delta[k]$ for $k \geq 0$ with $h[-1] = h[-2] = 0$. Explicit evaluation gives $h[0] = 1$ and $h[1] = 2\frac12$. From this point on, we have a one-sided recurrence. The standard theory of generating functions applies, and thus we know the general solution will be of the form $h[k] = c_1 \lambda_1^k + c_2 \lambda_2^k$ for $k \geq 0$ where $\lambda_1$ and $\lambda_2$ are the roots of the characteristic polynomial $\lambda^2 - 2\frac12 \lambda + 1$, namely, $\lambda_1 = \frac12$ and $\lambda_2 = 2$. The choice $c_1 = -\frac13$ and $c_2 = 1\frac13$ satisfy the conditions $h[0] = 1$ and $h[1] = 2\frac12$. Thus the impulse response of the associated LTI system is
\begin{equation} \label{eq:impresp}
  h[k] = \frac{2^{2k+2} - 1}{3 \times 2^k} u[k].
\end{equation}
This sequence is unbounded and does not have a discrete-time Fourier transform. How can we reconcile this with the frequency response $H(e^{\jmath 2 \pi f})$ found earlier?

Factoring $H(z)$ as
\begin{align}
  H(z) &= \frac1{(z^{-1}-\frac12)(z^{-1}-2)} \\
    &= \frac{1 \frac13}{1-2z^{-1}} - \frac{\frac13}{1-\frac12 z^{-1}}
\end{align}
shows that $H(z)$ has a pole at $z=\frac12$ and at $z=2$. We can therefore obtain three separate impulse responses by choosing the ROC to be $|z| < \frac12$, $\frac12 < |z| < 2$ and $2 < |z|$. By Fact 5, the choice $2 < |z|$ must yield the impulse response \eqref{eq:impresp}.

The frequency response $H(e^{\jmath 2 \pi f})$ is found by evaluating $H(z)$ on the unit circle $z = e^{\jmath 2 \pi f}$. Since $|e^{\jmath 2 \pi f}| = 1$, the ROC $2 < |z|$ is not applicable. Instead, we are concerned with the ROC $\frac12 < |z| < 2$ as this includes $|z| = 1$. Explicitly, this means expanding $H(z)$ as follows.
\begin{align}
  (1-\frac12 z^{-1})^{-1} &= \sum_{k=0}^\infty 2^{-k} z^{-k} \\
  (z^{-1} - \frac12)^{-1} &= z / (1-\frac12 z)
    = 2 \sum_{k=1}^\infty 2^{-k} z^k \\
  H(z) &= -1\frac13 \sum_{k=1}^\infty 2^{-k} z^k - \frac13 \sum_{k=0}^\infty 2^{-k} z^{-k}.
\end{align}
Therefore, the impulse response whose frequency response is $H(e^{\jmath 2 \pi f})$ is
\begin{equation}
  \tilde h[k] = \begin{cases}
    -\frac13 \times 2^{-k},&\quad k \geq 0, \\
    -1\frac13 \times 2^k,&\quad k < 0.
  \end{cases}
\end{equation}
This impulse response decays to zero at both ends, as $k \rightarrow \infty$ and as $k \rightarrow -\infty$.

To summarise, there are an infinite number of ``impulse responses'' of \eqref{eq:ar}, where by impulse response we mean any $\tilde h$ for which $(\delta[k],\tilde h[k])$ is a solution pair of \eqref{eq:ar}. Different impulse responses will in general have different frequency responses. When we use $H(e^{\jmath 2 \pi f})$ as the frequency response, we are implicitly using the impulse response $\tilde h$ obtained from $H(z)$ by using the ROC that contains the unit circle $|z|=1$. Provided all the poles are within the unit circle (implying that the system is stable) then this impulse response will be the same as that of the associated LTI system.

\section{What we have learned}

The general recursive filter \eqref{eq:ar} has an infinite number of possible outputs for any given input. By choosing an appropriate initialisation we can limit this to a single output. By choosing the initialisation to be $y[-N] = \cdots = y[-1] = 0$, we obtain a linear input-output mapping, but one that is not time invariant. We only really care about this filter for causal signals, hence we ask whether there is an LTI system that acts the same way as our filter on causal signals. It turns out there is, and we call it the associated LTI system. We can now study our filter by rigorously applying LTI theory to this associated LTI system.

The $z$-transform is essentially a generating-function approach to solving recurrence relations but because we are interested in doubly-infinite sequences we cannot work formally and must instead work with analytic functions. Multiple power series can be associated with the same analytic function $Y(z)$ if it is possible to express $Y(z)$ as $Y_0(z) + Y_\infty(z)$ in more than one way, where $Y_0(z)$ is analytic about 0 and $Y_\infty(z)$ is analytic about $\infty$. Moreover, $Y(z) = H(z) X(z)$ only makes sense for values of $z$ for which all three functions are defined. This is why it is necessary to keep track of the Region of Convergence. Fact 5 links the associated LTI system with the $z$-transform approach.

\appendix

Throughout, we assume infinite sums exist and are sufficiently well behaved (e.g., absolutely convergent) to permit the manipulations we apply.

\noindent\textbf{Proof of Fact 1:}
Let $\tilde y$ satisfy $\tilde y[k] = \sum_{i=-\infty}^\infty h[i] x[k-i]$. (Although $h[i]=0$ for $i<0$ it is simpler to keep the sum doubly infinite.) Then
\begin{align}
  \sum_{j=1}^N \alpha_j \tilde y[k-j]
    &= \sum_{j=1}^N \sum_{i=-\infty}^\infty \alpha_j h[i] x[k-j-i] \\
    &= \sum_{j=1}^N \sum_{i=-\infty}^\infty \alpha_j h[i-j] x[k-j-(i-j)] \\
    &= \sum_{i=-\infty}^\infty \sum_{j=1}^N \alpha_j h[i-j] x[k-i].
\end{align}
Now, by definition, $h[k]$ satisfies $h[k] = \sum_{j=1}^N \alpha_j h[k-j] + \delta[k]$. Substituting this into the previous calculation yields
\begin{align}
  \sum_{j=1}^N \alpha_j \tilde y[k-j]
    &= \sum_{i=-\infty}^\infty \left( h[i] - \delta[i] \right) x[k-i] \\
    &= \tilde y[k] - x[k].
\end{align}
Therefore, $\tilde y[k] = \sum_{j=1}^N \alpha_j \tilde y[k-j] + x[k]$, showing $\tilde y[k]$ is a valid output of the recursive filter \eqref{eq:ar}.

\bibliographystyle{IEEEtran}
\bibliography{AR-tutorial}

\vfill

\end{document}